\documentclass{elsart}

\def\cuscn{$\kappa$-(BEDT-TTF)$_2$Cu(NCS)$_2$}
\def\khg{$\alpha$-(BEDT-TTF)$_2$KHg(SCN)$_4$}
\def\tlhg{$\alpha$-(BEDT-TTF)$_2$TlHg(SCN)$_4$}
\def\mhg{$\alpha$-(BEDT-TTF)$_2M$Hg(SCN)$_4$}
\def\nh4{$\alpha$-(BEDT-TTF)$_2$NH$_4$Hg(SCN)$_4$}

\def\per{(Per)$_2M$(mnt)$_2$}
\def\pera{(Per)$_2$Au(mnt)$_2$}

\usepackage{graphicx}
\usepackage{amssymb}

\begin{document}

\begin{frontmatter}
\title{Recent high-magnetic-field
studies of unusual groundstates in 
quasi-two-dimensional crystalline organic metals
and superconductors}

\author[label1]{J. Singleton, N. Harrison, R. McDonald, P.A. Goddard}
\author[label2]{A. Bangura, A. Coldea}
\author[label3]{L. K. Montgomery}
\author[label4]{X. Chi}

\address[label1]{National High Magnetic Field 
Laboratory, LANL, MS-E536, Los Alamos, New Mexico 87545, USA}
\address[label2]{Department of Physics, University of Oxford, 
Clarendon Laboratory, Parks Road, Oxford OX1 3PU, UK}
\address[label3]{Department of Chemistry, Indiana University, 
Bloomington, Indiana 47405, USA}
\address[label4]{Bell Laboratories, Lucent Technologies, 
600~Mountain Avenue, Murray Hill, New Jersey 07974, USA}

\begin{abstract}
After a brief introduction to crystalline 
organic superconductors and metals, 
we shall describe two recently-observed exotic phases that occur 
only in high magnetic fields. 
The first involves measurements of the non-linear 
electrical resistance of single crystals of the 
charge-density-wave (CDW) system (Per)$_2$Au(mnt)$_2$ in static 
magnetic fields of up to 45~T and temperatures as low as 25~mK.
The presence of a fully gapped CDW state with typical 
CDW electrodynamics at fields higher that the 
Pauli paramagnetic limit of 34~T suggests the existence of a 
modulated CDW phase analogous to the 
Fulde-Ferrell-Larkin-Ovchinnikov state. Secondly, 
measurements of the Hall potential of 
single crystals of $\alpha$-(BEDT-TTF)$_2$KHg(SCN)$_4$, made 
using a variant of the Corbino geometry in quasistatic magnetic 
fields, show persistent current effects that are similar to 
those observed in conventional superconductors. 
The longevity of the currents, large Hall angle,
flux quantization and 
confinement of the reactive component of the Hall potential
to the edge of the sample are all consistent with the 
realization of a new state of matter in CDW systems with 
significant orbital quantization effects in strong magnetic fields. 
\end{abstract}

\end{frontmatter}

\section{Introduction}
\label{intro}
Quasi-two-dimensional crystalline organic metals and 
superconductors are very flexible systems in the study 
of many-body effects and unusual mechanisms for 
superconductivity~\cite{r0a,r0b,r1,r1a,r1b,r2,r3}.
Their ``soft'' lattices enable one to use 
relatively low pressures 
to tune the same material through a variety of 
low-temperature groundstates, for example from Mott 
insulator via intermingled 
antiferromagnetic and superconducting 
states to unusual superconductor~\cite{r1a,r2,r3}. 
Pressure also provides a sensitive means of varying the 
electron-phonon and electron-electron interactions, 
allowing their influence on the superconducting 
groundstate to be mapped~\cite{r1,r1a,r4}. 
The self-organising 
tendencies of organic molecules means that organic metals 
and superconductors are often rather clean and 
well-ordered systems, enabling the Fermi-surface 
topology to be measured in very great detail
using modest magnetic fields~\cite{r1,r5};
such information can then be used as input
parameters for theoretical models~\cite{r1}.
And yet the same organic molecules can adopt a variety 
of configurations, leading to ``glassy''
structural transitions and mixed phases in otherwise very pure 
systems~\cite{r1a,r6,r7}; these states may be important 
precursors to the superconductivity in such cases~\cite{r7}. 

Intriguingly, there seem to be at least two (or possibly three)
distinct mechanisms for superconductivity~\cite{r1,r8,r9,r22}
in the quasi-two-dimensional organic conductors.
The first applies to half-filled-band
layered charge-transfer salts, such as the
$\kappa-$, $\beta-$ and $\beta'-$ 
packing arrangements of salts
of the form (BEDT-TTF)$_2$X,
where X is an anion molecule;
the superconductivity appears to be 
mediated by electron correlations/ antiferromagnetic 
fluctuations~\cite{r1,r1a,r1b}.
The second mechanism applies to 
{\it e.g.} the $\beta''$ phase BEDT-TTF salts~\cite{r1a};
it appears to depend on the proximity of a metallic phase to 
charge order~\cite{r7,r8,r9}.
Finally, there may be some instances 
of BCS-like phonon-mediated
superconductivity~\cite{r22}.

\begin{figure}[htbp]
\centering
\includegraphics[width=9cm]{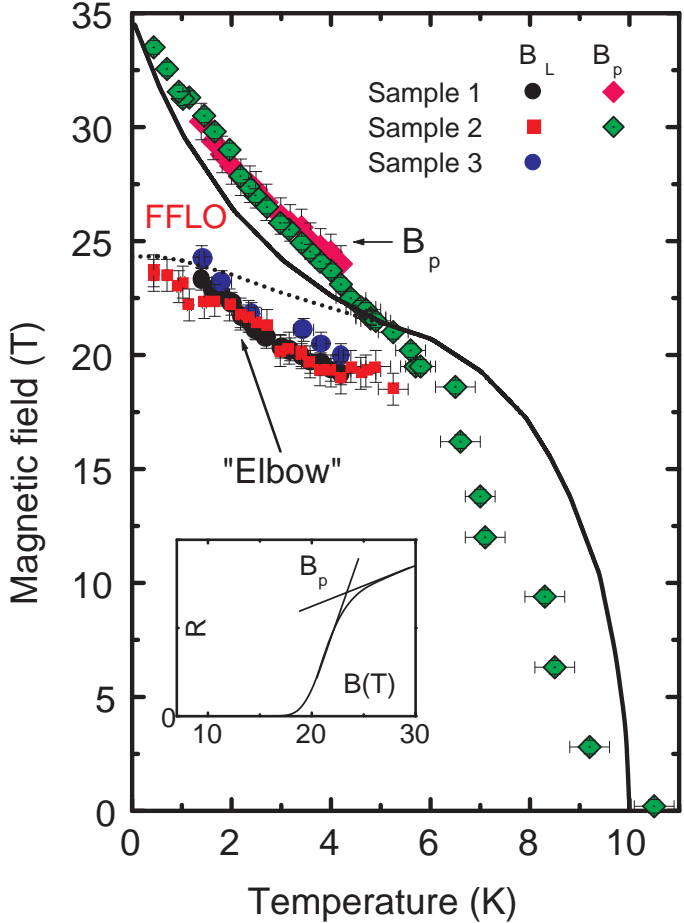}
\caption{Observation of the Fulde-Ferrell-Larkin-Ovchinnikov
(FFLO) phase in \cuscn ~\cite{r11}. The points labelled
$B_{\rm P}$ denote the resistive upper critical field
for two different samples (see inset for an 
illustrative example).
The $B_{\rm L}$ points, denoting
the phase boundary between the mixed
phase and FFLO state were deduced using simultaneous
MHz differential susceptibility measurements;
the change in vortex stiffness that accompanies
entry into the FFLO state causes an ``elbow''
in the field-dependent susceptibility.
``Sample 3'' is a measurement on a third sample
under different conditions of electric field;
consistency of the phase boundaries for the
three samples shows that the effect is not
due to artefacts of vortex pinning. The
curves are a theoretical model
due to Shimahara for the 
upper critical field and FFLO
(see Ref.~\cite{r11} for details).}
\label{fig1}
\end{figure}

In addition to their invaluable role in mapping the
bandstructure, high magnetic fields
allow one to tune some of the organic
conductors into some new and intriguing phases;
examples include field-induced superconductivity~\cite{r10}
and exotic states such as the Fulde-Ferrell-Larkin-Ovchinnikov 
(FFLO) phase~\cite{r11} (Fig.~\ref{fig1}). 
In the remainder of this paper, 
we shall describe two recent observations of 
field-induced phases in crystalline organic metals, 
one of which is related to the FFLO but which results in 
an insulating state, and the other of which exhibits 
properties reminiscent of those of a superconductor, 
but via an entirely different mechanism.

\section{Charge-density waves at fields above 
the Pauli paramagnetic limit}
\label{pauli}
Intense magnetic fields ($B$)
impose severe constraints 
on spin-singlet paired electron states. 
Superconductivity is one example of a 
groundstate where this is true, although 
orbital diamagnetic effects usually destroy 
superconductivity at lower 
magnetic fields than does 
spin-splitting~\cite{r11,r12}. 
Charge-density wave (CDW) systems, by comparison, 
are mostly free from orbital effects~\cite{r13}, 
and so can only be destroyed by coupling 
$B$ directly to the electron spin. 
While most CDW systems have gaps that are 
too large to be destroyed in laboratory-accessible 
fields~\cite{r13}, several new compounds 
have been identified within the last decade 
that have gaps ($2\Psi_0$) as low as a few meV, 
bringing them within range of the static magnetic 
fields at the National High Magnetic Field Laboratory.

As we shall discuss in Section~\ref{three},
\mhg (where $M = $K, Tl or Rb; $2\Psi_0 \sim 4$~meV) 
is one example that has been extensively studied~\cite{r14}. 
However, it has a complicated phase diagram in a 
magnetic field owing to the imperfect nature of 
the nesting~\cite{r15}; closed orbits exist 
after the Fermi-surface reconstruction 
which become subject to Landau quantization in a 
magnetic field~\cite{r16}, potentially modifying 
the groundstate. By contrast, 
\per ~(where M = Pt and Au) appears to be 
fully gapped~\cite{r17}.
However, the existence of spin $\frac{1}{2}$
moments on the Pt sites makes only the $M =$Au 
system a pristine example of a small-gap, fully 
dielectric CDW material. 

Measurements of the CDW transition temperature $T_{\rm P}$
(where the subscript ``P'' stands for ``Peierls'')
in \pera ~
($T_{\rm P} = 11$~K at $B=0$) as a function of 
$B$ indicate that it is suppressed in a 
predictable fashion~\cite{r18}, 
allowing a Pauli paramagnetic limit
of $B_{\rm P} \approx 37$~T to be inferred.
However, the closure of the CDW gap with field is 
in fact considerably more subtle~\cite{r19};
a finite transfer integral $t_a$ perpendicular to the
nesting vector produces a situation analogous to
that in an indirect-gap semiconductor, where
the minimum energy of the empty states above the chemical
potential $\mu$ is displaced in $k$-space
from the maximum-energy occupied states
below $\mu$~\cite{r19}.
Consequently, Landau quantization
of the states above and below $\mu$
is possible, leading to a
thermodynamic energy gap $E_{\rm g}(B,T)$ of
the form~\cite{r19}
\begin{equation}
\label{gapfield}
E_{\rm g}(B,T)=2\Psi(T)-4t_a-g\mu_{\rm B}B
+\gamma\hbar\omega_{\rm c}.
\end{equation}
Here, $\Psi(T)$ is the temperature-dependent
CDW order parameter ($\Psi(T) \rightarrow \Psi_0$ 
as $T \rightarrow 0$), 
$\omega_{\rm c}$ is a characteristic cyclotron frequency
in the limit $B\rightarrow 0$, and $\gamma$
is a nonparabolicity factor;
$g\approx 2$ is the Land\'{e} g-factor
and $\mu_{\rm B}$ is the Bohr magneton.
Note that the Landau quantization
{\it competes} with 
Zeeman splitting; however, at sufficiently
high $B$ it becomes impossible to sustain closed
orbits, leading to a straightforward dominance of the
Zeeman term~\cite{r19}.

\begin{figure}[htbp]
\centering
\includegraphics[width=6cm]{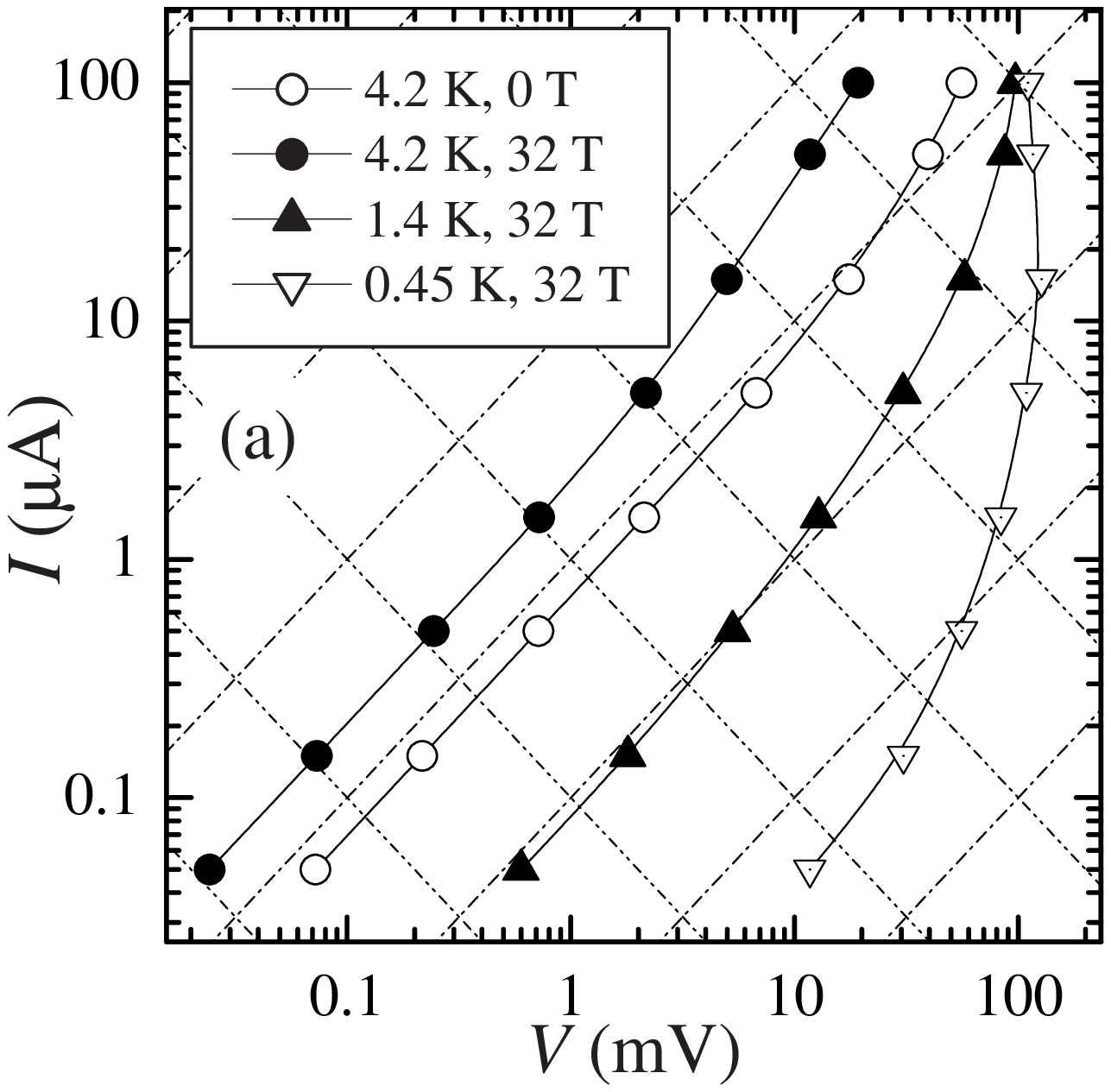}
\includegraphics[width=6cm]{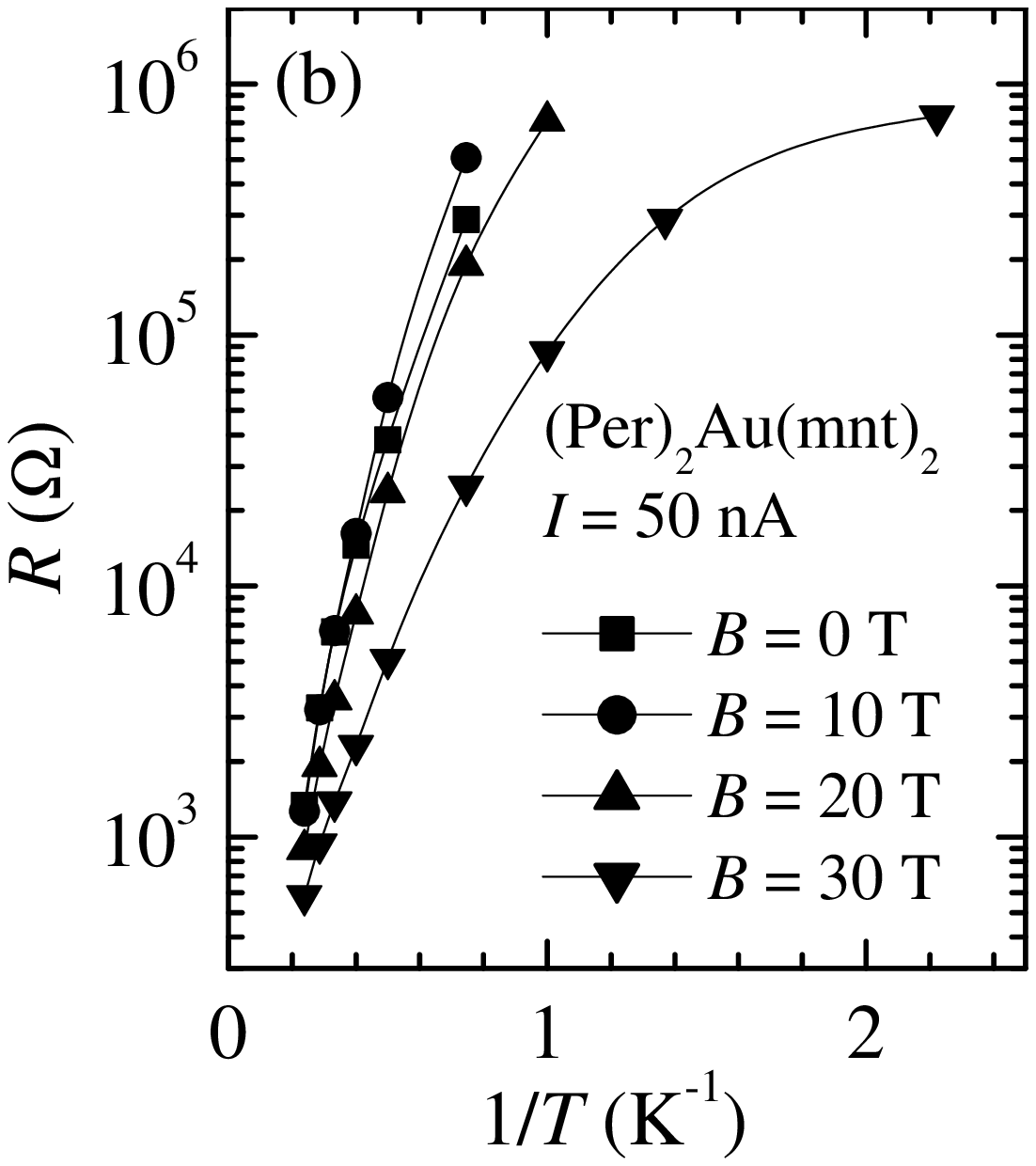}
\caption{(a)~Non-linear current-versus-voltage 
characteristics of \pera ~
plotted on a logarithmic scale for various
temperatures and fields (see inset key).
The negative-slope diagonal lines are contours 
of constant power and the 
positive-slope diagonal lines are contours of 
constant resistance, providing a guide 
as to when the sample's behavior is dominated by ohmic,
thermally-activated conduction rather than sliding.
(b)~Arrhenius plots of resistance $R$ (
$\propto \rho_{yy}$, logarithmic scale)
versus $1/T$ with $I=$~50~nA 
for \pera ~at several different
$B$ (after Ref.~\cite{r19}).}
\label{fig2}
\end{figure}

Another subtlety in assessing the field-dependent 
thermodynamic gap (and hence the Pauli
paramagnetic limit) in \pera ~stems from
the complicated nature of the low-temperature 
electrical conductivity, which contains contributions
from both the sliding collective mode of the CDW
and thermal excitation across the gap 
(see Fig.~\ref{fig2}(a))~\cite{r19,r20}.
This leads to a measured resistivity 
$\rho_{yy}\approx (\sigma_T+j_y/{E}_{\rm t})^{-1}$,
where $\sigma_T$ is the conductivity
due to thermal excitation across the
gap and $j_y/{E}_{\rm t}$ is the 
contribution from the collective mode,
$j_y$ being the current density and ${E}_{\rm t}$
the threshold field. 
Now Eq.~\ref{gapfield}
contains $\Psi$, which is $T$-dependent;
moreover, ${E}_{\rm t}$ may also
depend on $T$.
Thus, Arrhenius plots are in general curved 
(see Fig.~\ref{fig2}(b)),
with a slope 
\begin{equation}\label{slope}
\frac{\partial\ln\rho_{yy}}{\partial (1/T)} \approx 
\frac{\frac{1}{2k_{\rm B}}(E_{\rm g}-T\frac{\partial E_{\rm  g}}{\partial T})
-\frac{j_yT^2}{\sigma_T{E}_{\rm t}^2}    
\frac{\partial {E}_{\rm t}}{\partial T}}
{1+\frac{j_y}{\sigma_T {E}_{\rm t}}}.
\end{equation}
With appropriate choice of temperature and bias regimes,
it is possible to make a reliable estimate
of $E_{\rm g}$ from plots such as those
in Fig.~\ref{fig2}(b). On the other
hand, it can be shown~\cite{r20a} that
poorly-chosen experimental conditions
(e.g. those of Ref.~\cite{r18})
can easily lead to errors in the
size of the derived gap.

Once accurate values of $E_{\rm g}(B)$
have been obtained, the method of
Ref.~\cite{r19} can be used to fit Eq.~\ref{gapfield}
by adjusting the parameters
$2\Psi_0+4t_a$, $4t_a$
and $v_{\rm F}$, where $v_{\rm F}$ is the Fermi velocity
in the metallic state;
in the CDW state it is used to parameterise
the quasiparticle dispersion.  
A good fit is obtained
using
$t_a=$~0.20~$\pm$~0.01~meV,
$v_{\rm F}=$~(1.70~$\pm$~0.05)~$\times 10^5$~ms$^{-1}$
and $2\Psi_0 =$~4.02~$\pm$~0.04~meV~\cite{r19}.
These parameters correspond to 
$E_{\rm g}=$~3.21~$\pm$~0.07~meV 
at $B=0$, $T=0$; the derived
transfer integrals
$t_a$ and $t_b~(\approx$~188~meV) are
in good agreement with theory~\cite{r20b}
and thermopower data~\cite{r17}.
(Note that these band parameters
exclude the possibility
of field-induced CDW (FICDW) states of the
kind proposed in Ref.~\cite{r20c} in \per ~salts.)

\begin{figure}[htbp]
\centering
\includegraphics[width=9cm]{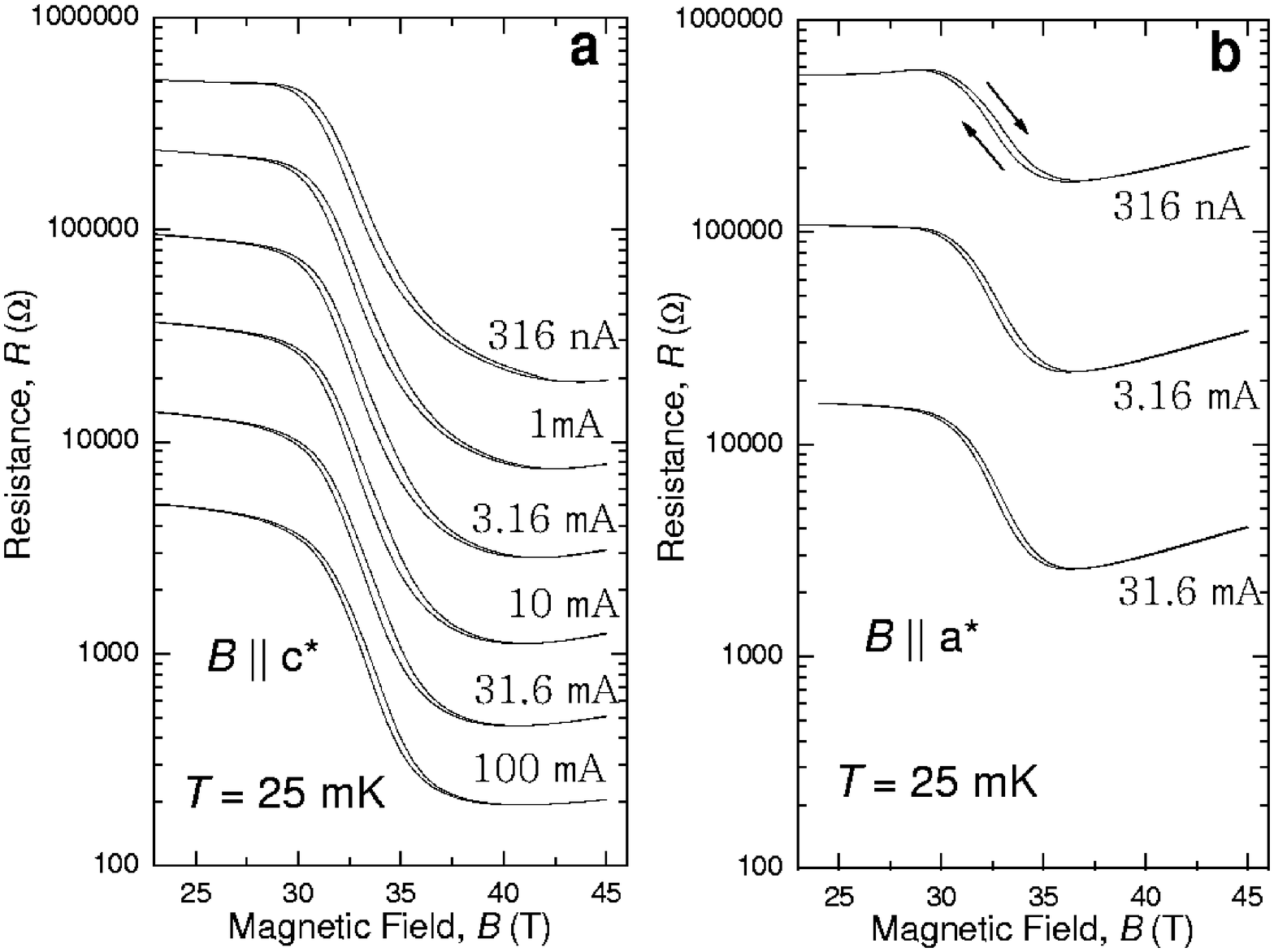}
\includegraphics[width=4.5cm]{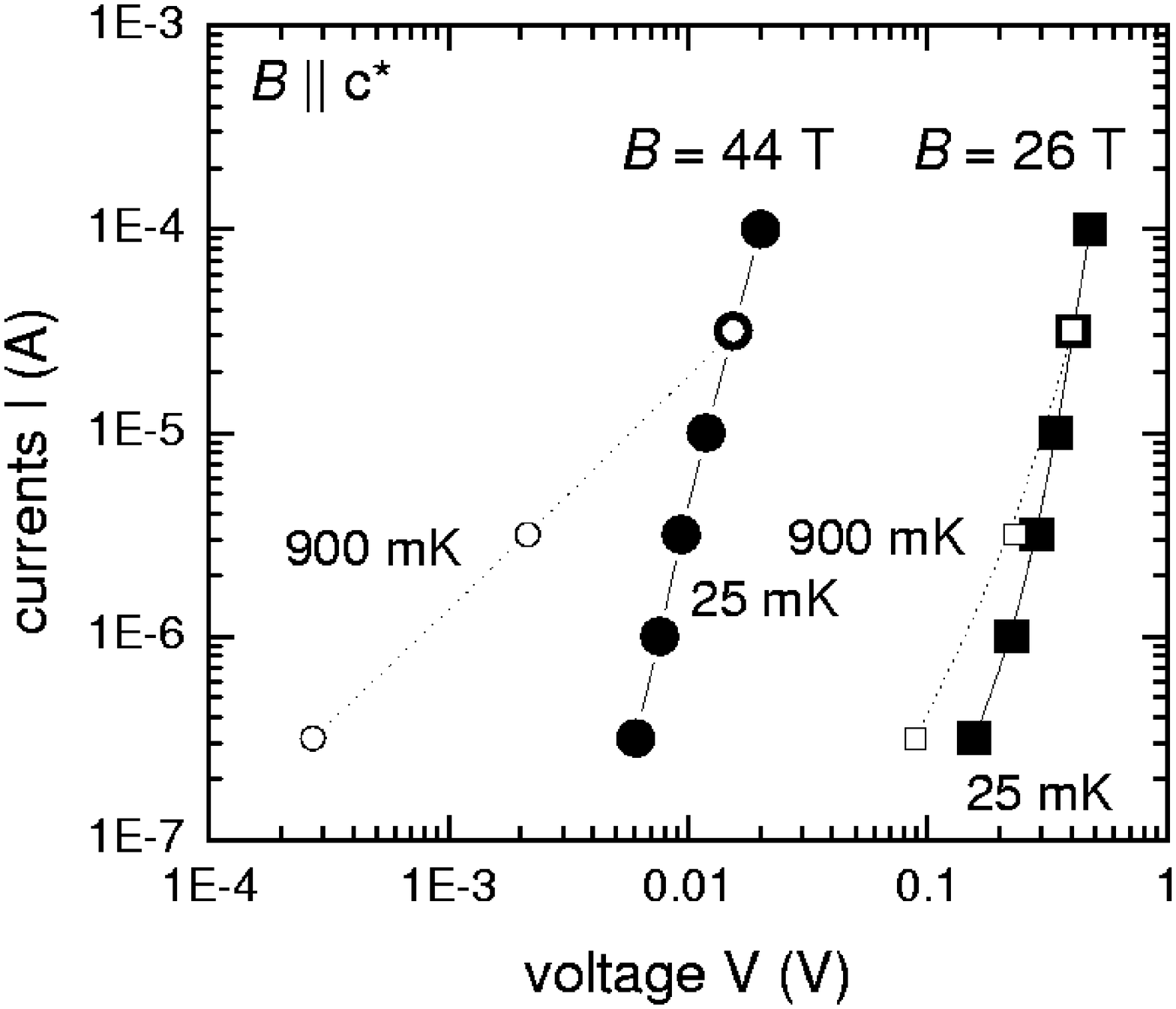}
\caption{Left: resistance of a single crystal of 
(Per)$_2$Au(mnt)$_2$
measured at 25~mK for
fields between 23 and 45~T, 
for two different orientations $c^\ast$ (a) and
$a^\ast$ (b) of $B$ perpendicular to its 
long axis $b$, at several different
applied currents.  
The lowest resistance for a given current occurs
for $B$ parallel to $c^\ast$, 
which is perpendicular to $a^\ast$. 
The dependence of
the resistance on current signals non-ohmic behaviour.
Right:~non-linear current-versus-voltage characteristic of
(Per)$_2$Au(mnt)$_2$ plotted on a log-log scale, 
for magnetic
fields (26 and 44~T) above (circles) and below (squares) $B_{\rm P}$. 
Filled symbols connected by solid 
lines represent data taken at 25~mK
while open symbols connected by 
dotted lines represent data taken at
900~mK. (After Ref.~\cite{r20}.)}
\label{fig3}
\end{figure}

Armed with the band parameters,
one obtains a reliable estimate
of the Pauli paramagnetic limit; 
$B_{\rm P}=(\Delta_0+2t_a)/\sqrt{2}gs\mu_{\rm B}\approx 30$~T.
This corresponds to a sharp drop in
measured resistance $R$ ($\propto \rho_{yy}$),
as shown in Fig.~\ref{fig3} (left side),
which displays data recorded at
$T=25$~mK.
At such temperatures, there are very
few thermally-activated 
quasiparticles indeed,
leaving only the CDW collective mode to conduct;
this gives rise to a $R$ 
in Fig.~\ref{fig3} 
that is strongly dependent on current.
On passing through $B_{\rm P}\approx 30$~T,
$R$ drops very sharply,
and there is also hysteresis between up- and
down-sweeps of the field.
The latter effect could be the consequence
of a first-order phase transition on reaching
$B_{\rm P}$, compounded by CDW pinning effects.  

However, the most interesting observation about
Fig.~\ref{fig3} is the fact that 
the strongly non-linear $I-V$ characteristics
persist at fields well above $B_{\rm P}$,
as can be seen from both
$R$ data and $I-V$
plots (right-hand side of
Fig.~\ref{fig3}).
One can conclude from these data that it is 
only the threshold voltage for depinning the CDW that 
changes at $B_{\rm P}$. This is probably a consequence of the 
CDW becoming incommensurate or of the order parameter 
of the charge modulation becoming considerably 
weakened~\cite{r13,r20}. Evidence for the latter is obtained 
by repeating the $I-V$ measurements at slightly higher 
temperatures of 900~mK (Fig.~\ref{fig3}, right side). 
This temperature is sufficient 
to restore Ohmic behaviour for $B > B_{\rm P}$, 
suggesting that a 
reduced gap for $B > B_{\rm P}$ 
allows quasiparticles to be more 
easily excited. 

There have been some interesting
proposals for FICDW phases (e.g. Ref.~\cite{r20c}).
These mechanisms would not cause the 
system to be completely gapped, 
but would instead lead to a closed orbit for 
one of the spins which would then have a Landau gap 
at the chemical potential. Such a situation leads to 
the quantum Hall effect, and a metallic  behaviour 
of longitudinal resistivivity~\cite{r21}. 
Current would then be able to flow without the 
CDW having to be depinned.
However, these effects are plainly absent
from the data of Fig.~\ref{fig3};
the continuation of the non-linear CDW electrodynamics 
for $B > B_{\rm P}$  shows that
metallic behaviour is not regained.
Instead, both spin 
components of the Fermi surface are 
almost certainly gapped independently
with differing nesting vectors,
leading to an exotic CDW phase 
that has some analogies with the
FFLO state of superconductors. 
The presence of two distinct, spin-polarized
CDWs with different periodicities 
will furnish separate spin and charge modulations 
that could in principle be detected using 
a diffraction experiment~\cite{r20}.
\section{A new quantum fluid in strong magnetic fields with 
orbital flux quantization}
\label{three}
\khg ~is undoubtedly one of the most 
intriguing of BEDT-TTF-based charge-transfer 
salts~\cite{r14,r15,r16}. 
Like many other such materials, it possesses 
both two-dimensional (2D) and one-dimensional (1D) 
Fermi-surface (FS) sections. 
However, the 1D sheets are unstable at 
low temperatures, causing a structural phase 
transformation below $T_{\rm P} = 8$~K into a CDW state~\cite{r23}.
Imperfect nesting combined with the continued existence 
of the 2D hole FS pocket gives rise to 
complicated magnetoresistance and 
unusual quantum-oscillation 
spectra at low magnetic fields and low temperatures~\cite{r22}. 
At high fields, the CDW undergoes a number of 
transformations into new phases, many of which 
have been suggested to be 
field-induced CDW phases~\cite{r15}.

\begin{figure}[htbp]
\centering
\includegraphics[width=8cm]{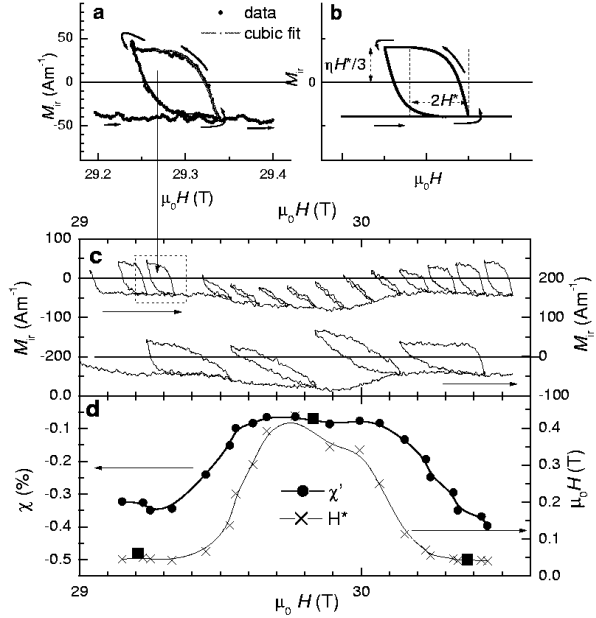}
\caption{Evidence for persistent currents
and critical state behaviour 
in \khg.  ~(a) An
example of a loop in the non-equilibrium 
component of the magnetization $\Delta M$ measured 
as the field is swept up to
29.33~T then momentarily down to 29.24~T 
before being resumed.  Arrows
indicate the change in the locus of 
$\Delta M$ versus $H$.  The
solid lines show the results of 
fits of $\Delta M$ to a Bean-type
model of the critical state.
(b) A model hysteresis loop calculated
according to 
the Bean model for a long cylinder with its 
axis parallel to $H$, 
showing the theoretical saturation value of
$\Delta M_z$ and the reversal state.  
(c) A series of $\Delta M$
versus $H$ loops like that in (a) measured over 
an extended interval
of field (top), and (bottom), 
a similar measurement but where the
field interval over which the sweep 
direction is reversed is increased.  
Arrows indicate the appropriate axes.  (d) Estimates of
$\Delta\chi$ from fits of a Bean model 
to the many loops shown in
(c) (full circles)
Estimates of the coercion field $H^\ast$ (x-symbols) are also shown.
Arrows indicate the appropriate axes and square points
are theoretical simulations. (After Ref.~\cite{r16}.)}
\label{fig4}
\end{figure}

Undoubtedly the most exotic aspect of this material is 
its transformation into a new state of matter 
above a characteristic field $B_{\rm k} = 23$~T 
(known as the ``kink'' transition); 
$B_{\rm k}$ is now known to correspond to the 
CDW Pauli paramagnetic limit~\cite{r16,r22}. 
Such a regime is reached in \khg ~owing to the 
unusually low value  of $T_{\rm P}$~\cite{r16}. 
At fields higher than $B_{\rm k}$, Zeeman splitting 
of the energy bands makes a conventional CDW 
ground state energetically unfavourable~\cite{r24}, 
possibly yielding a novel modulated CDW state like 
that proposed for \pera ~(see Section~2
and Refs.~\cite{r19,r20}). 
In \khg ~this state is especially unusual due to 
the existence  of the 2D pocket, which appears 
to be ungapped by CDW formation~\cite{r16}. The CDW 
and 2D hole pocket screen each other, 
with pinning of the CDW then enabling a 
non-equilibrium distribution of orbital 
magnetization throughout the bulk~\cite{r16}. 
Consequently,
such a state exhibits a critical state analogous 
to that in type II superconductors, as shown in
Fig.~\ref{fig4}. 

This same non-equilibrium distribution also 
puts the CDW under considerable tension or compression, 
rather like a spring. Orbital quantization 
of the 2D pocket enables the magnetic field 
to be directly coupled to this ``spring'' via the 
variation of the chemical potential~\cite{r16,r26}. 
The magnetic field therefore provides a means 
to drive the compression and tension of the CDW 
and subsequent development of non-equilibrium 
persistent currents~\cite{r16,r26}.

Attempts to drive a bulk current though 
\khg ~at low temperatures and $B > B_{\rm k}$, 
either inductively or by way of external contacts, 
causes the CDW to undergo compression or tension~\cite{r16}. 
The build-up of potential energy in 
response to the current can be particularly 
strong at integral filling factors of the 
2D pocket~\cite{r16}, giving rise to an effective 
force that is orthogonal to the current and 
the magnetic field and which varies approximately 
linearly across the width of the sample. 
Thus, there is a considerable energetic advantage 
in forcing the current towards the surface. 
Such a process protects the normal quasiparticles 
from inelastic scattering events, enabling the 
transport to become ballistic~\cite{r16}. 
The electrodynamics of the 2D hole gas can then 
become significantly different than those of 
a regular dissipative material, allowing 
charge to accumulate inside the bulk 
that gives rise to an electric field 
orthogonal to the current $j$. 

The electrodynamics can be summarized as follows~\cite{r16}:
\begin{equation}
\frac{1}{2 \omega_{\rm c0}}\nabla^2 V + \Delta B = 0;
\end{equation}
\begin{equation}
\Delta B = -\frac{\mu_0 \rho_{\rm 2D}}{B_0}V,
\end{equation}
where $V$ is the electrostatic potential,
$\Delta B$ is the perturbation in background magnetic 
field $B_0$ due to a current, $\omega_{\rm c0}$ 
is the unperturbed cyclotron frequency 
and $\rho_{\rm 2D}$ is the charge density 
of the 2D pocket~\cite{r16}. The first of 
these equations results from the effect 
of a charge build-up on the cyclotron motion 
and Landau level degeneracy whereas the 
second is simply the Hall effect in the absence 
of scattering. On combining these Equations, 
we obtain
\begin{equation}
\lambda^2 \nabla^2 V = V,
\label{eq3}
\end{equation}
where $\lambda= (m^*/2 \mu_0 e \rho_{\rm 2D})^{1/2}$
is analogous to the London penetration 
depth in superconductors~\cite{r16};
here $m^*$ is the 2D quasiparticle effective mass.
Eq.~\ref{eq3} describes an exponential 
screening of electric fields and currents 
from within the bulk~\cite{r16}; it also suggests that 
an ideal Hall effect should occur in transport experiments, 
whereby a Hall angle 
$\Theta = \tan^{-1}(\rho_{xy}/\rho_{xx}) \rightarrow 90^{\circ}$
is observed.

Transport measurements made by applying contacts 
directly to samples of \tlhg ~(which has the same 
groundstate as \khg) 
have thus far proved inconclusive, 
owing mostly to the measured in-plane
resistivity being contaminated 
by contributions from the 
interlayer component~\cite{r27}. However,
the Corbino 
geometry provides an alternative means of 
investigating the Hall voltage.
In the case of \khg, ~this is done according 
to the method of Ref.~\cite{r28}. 
Because bulk single crystals of
\khg ~cannot be easily machined into 
arbitrary shapes, contacts must be placed 
between the geometric center of the circulating 
current path and the sample edge~\cite{r16,r28}.
\begin{figure}[htbp]
\centering
\includegraphics[width=8cm]{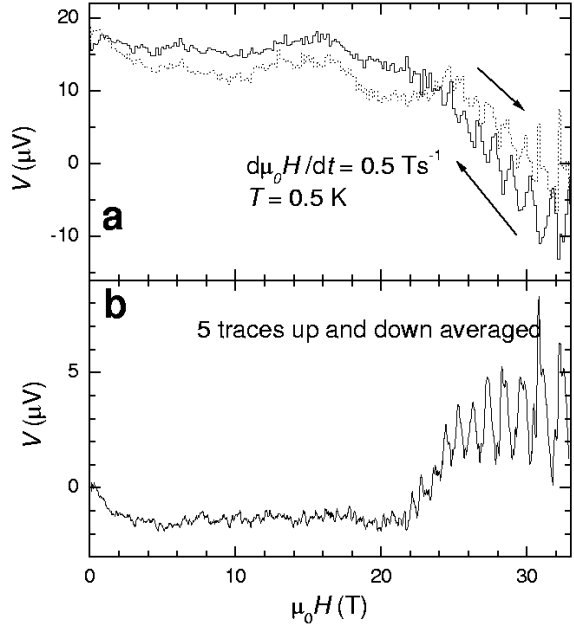}
\caption{An example of the Hall potential 
$V_{\rm H}$ of a \khg ~sampple in 
a slowly varying magnetic field
(sweep rate $\partial\mu_0H/\partial t=$ 0.5~T$^{-1}$; $T=0.50$~K).
(a) shows raw data for a single sweep, 
with arrows indicating the sweep direction.
No filtering has been applied. 
The background non oscillatory
component could be caused by variations
in the contact potentials with
magnetic field. (b) The difference 
$(V_{\rm H,up}-V_{\rm H,down})/2$
between rising- and falling-magnetic-field 
data averaged over 5 sweeps. (After Ref.~\cite{r16}.)}
\label{fig5}
\end{figure}

\begin{figure}[htbp]
\centering
\includegraphics[width=11cm]{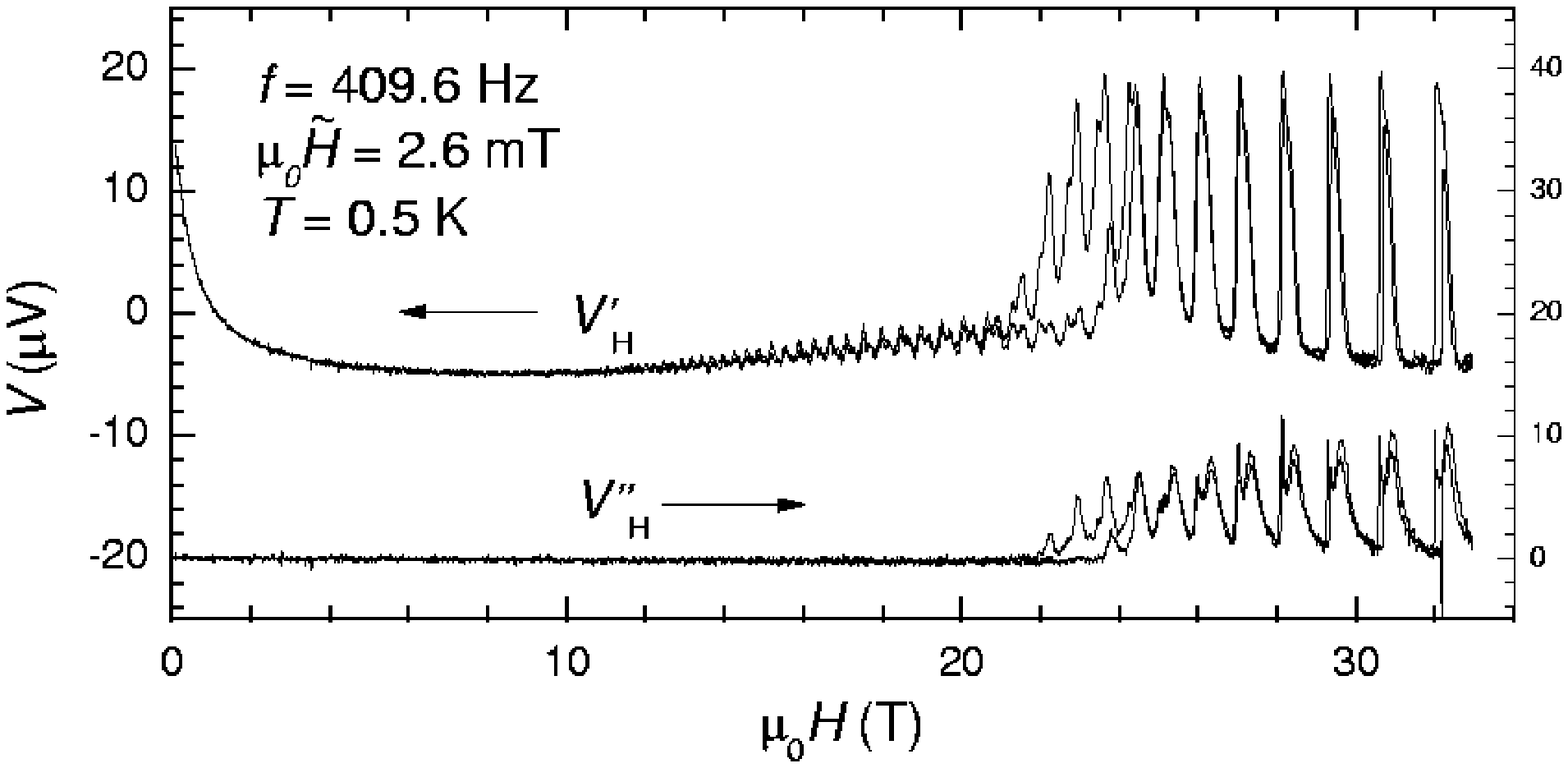}
\includegraphics[width=11cm]{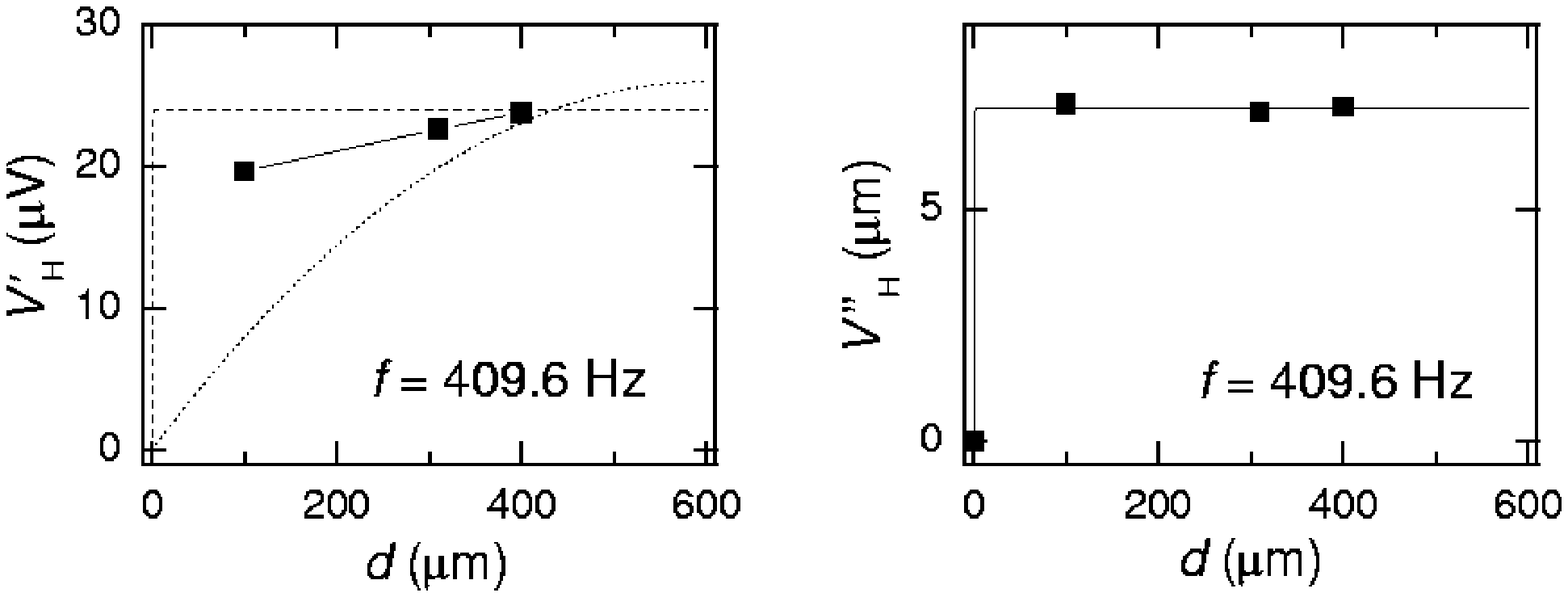}
\caption{Top: examples of the in-phase 
and quadrature Hall potentials
$V^\prime_{\rm H}$ and $V^{\prime\prime}_{\rm H}$
in \khg.
~An oscillatory
applied field with $\mu_0\tilde{H}\approx 2.6$~mT rms 
and $\omega/2\pi =409.6$~Hz was used; $T\sim$~0.5~K.
Arrows indicate the axis corresponding to each data set.
Bottom: the Hall potential difference 
between the edge and a voltage
probe situated a distance $d$ 
inside the upper surface where $d=$~100, 310 and 400~$\mu$m 
with
$T=$~0.5~K, $f=$~409.6~Hz and 
$\mu_0\tilde{H}=$~2.6~mT. (a) shows the
normal Hall voltage $V^\prime_{\rm H}$.  
The dotted line shows the
voltage distribution expected
with a very high conventional conductivity while the 
dashed line shows the voltage
distribution expected for an exponential variation
of the Hall potential, $V_{\rm H} \propto \exp(-d/\lambda)$.  
(b) shows the reactive Hall 
voltage $V^{\prime\prime}_{\rm H}$, with the 
solid line depicting $V_{\rm H} \propto \exp(-d/\lambda)$ 
with $\lambda\ll$~100~$\mu$m.
(After Ref.~\cite{r16}.)}
\label{fig6}
\end{figure}

This experimental scheme 
constitutes perhaps the simplest possible measurement 
that can made in a magnetic field~\cite{r16}. 
The two contacts are connected directly to the 
input of a voltmeter that detects the induced 
voltage as the field is slowly swept. 
Data are shown in Fig.~\ref{fig5}, in which 
the Hall voltage 
is given by 
\begin{equation}
V_{\rm H}= \frac{A \mu_0}{4 \pi} \frac{\partial H}{\partial t}
\tan \Theta,
\label{eq5}
\end{equation}
where $A \approx 1$~mm$^2$ is the cross-section 
of the sample. Eq.~\ref{eq5} implies that
$\Theta = 89.7^{\circ}$ at the highest fields, 
which is perhaps the highest value ever 
reported for a single-crystal sample. 
The greatest sensitivity is, however, obtained by 
using a small sinusoidal ac magnetic field 
superimposed on top of the background field, 
which enables the normal (i.e. in phase with the 
voltage induced in a conventional coil) 
and reactive components to be extracted~\cite{r16}. 
A reactive Hall voltage component $V''_{\rm H}$ indicates 
the existence of long-lived free currents, 
and is only observed at fields $B > B_{\rm k}$ (Fig~\ref{fig6}); 
this is expected by virtue of the fact that 
this new state of matter only exists in the 
high-magnetic-field regime. 

If the reactive current is confined
mostly within a distance $\lambda$ of the sample edge, 
the Hall voltage drop should also occur mostly within 
that distance~\cite{r16}. Fig.~\ref{fig6} shows 
that the reactive Hall voltage drop occurs 
within a distance that is too small to be 
detected using regular contacts applied using graphite 
paint. The normal component is also unusual, 
but shows a weak penetration of the normal 
in-phase Hall electric field into the bulk.

In summary, measurements of the Hall potential 
difference on single crystals of \khg ~provide 
compelling evidence for this existence of 
a new type of quantum fluid resulting 
from the mutual coupling of a 2D hole gas 
to a gapped CDW system~\cite{r16}. Such a 
large reactive Hall potential difference 
is not observed in conventional metals. 
In this quantum fluid, the build up of charge 
in the bulk within a distance $\lambda$ of the surface 
prevents the CDW from having to be coerced by a 
changing magnetic field into a non-equilibrium 
state, thereby conserving energy. 

Finally, it should be noted
that an analysis 
of the ratio of magnetic flux saved 
per $2\pi$ phason of the CDW 
(via trivial algebraic manipulation of the results
of Ref.~\cite{r16}) reveals an unusual 
virtual quantization of magnetic flux per phason 
of the form
\begin{equation}
\Phi_0=\frac{h}{2\nu e},
\end{equation}
where $\nu$ is the Landau-level
filling factor of the 2D hole system at integral Landau-level
filling factors. The flux quantization is
irrational at other filling factors.

\section*{Acknowledgements}
This work is funded by US Department of Energy 
(DoE) grant LDRD 20040326ER. Work at NHMFL 
is performed under the auspices of the 
National Science Foundation, DoE and the 
State of Florida.

\end{document}